\documentclass[lettersize,journal]{IEEEtran}
\usepackage{amsmath,amssymb,amsfonts}
\usepackage{algorithmic}
\usepackage{graphicx}
\usepackage{textcomp}

\usepackage{amsmath,amsfonts}
\usepackage{algorithmic}
\usepackage{algorithm}
\usepackage{array}
\usepackage[caption=false,font=normalsize,labelfont=sf,textfont=sf]{subfig}
\usepackage{textcomp}
\usepackage{stfloats}
\usepackage{url}
\usepackage{verbatim}
\usepackage{graphicx}
\usepackage{enumerate}
\usepackage{here}
\usepackage{mathtools}
\usepackage{multirow} 
\usepackage{makecell}  
\usepackage{cuted} 
\usepackage{stfloats} 
\usepackage{xcolor}
\usepackage{booktabs}

\usepackage[nocompress]{cite}

\usepackage[]{hyperref}
\hypersetup{
	colorlinks=true, 
	citecolor=blue,
	linkcolor=red,
	urlcolor=orange,
	citebordercolor=green,
	linkbordercolor=red,
	urlbordercolor=cyan,
}

\begin{document}
\bstctlcite{IEEEexample:BSTcontrol}
\newtheorem{rem}{Remark}

\title{
	Physics-Informed Direct Data-Driven Gain-Scheduled Control for Improving the Extrapolation Performance of Vehicle Lateral Control
}

\author{
	Shuichi Yahagi,
	Aoi Koyanagi,
	Ansei Yonezawa,\\
  Heisei Yonezawa,
  Hiroki Seto,
	Toshiyuki Sugimachi,
	and Itsuro Kajiwara
\thanks{
		Manuscript received xx xx, 20xx; revised xx x, 20xx. 

		Shuichi Yahagi (Corresponding author) and Toshiyuki Sugimachi are with the Department of Mechanical Engineering, Tokyo City University, 
		1-28-1 Tamazutsumi, Setagaya-ku, Tokyo, 158-8557, Japan
		(e-mail: yahagis@tcu.ac.jp).

		Aoi Koyanagi is with the Graduate School of Integrative Science and Engineering, Tokyo City University, 
		1-28-1 Tamazutsumi, Setagaya-ku, Tokyo, 158-8557, Japan.

    Ansei Yonezawa is with the Faculty of Information Science and Technology, Hokkaido University,
		N14, W9, Kita-ku, Sapporo 060-0814, Japan.
    
    Heisei Yonezawa and Itsuro Kajiwara are with the Division of Mechanical and Aerospace Engineering, Hokkaido University, N13, W8, Kita-ku, Sapporo, Hokkaido 060-8628, Japan.
    
    Hiroki Seto is with the 6th Research Department, ISUZU Advanced Engineering Center Ltd., 8 Tsutidana, Fujisawa-shi, Kanagawa 252-0881, Japan.
}
}

\maketitle

\begin{abstract}
This paper proposes a physics-informed direct data-driven control (PI-D3C) framework for designing a gain-scheduled proportional-integral controller for vehicle yaw-rate tracking. Vehicle lateral dynamics vary depending on vehicle velocity, making it difficult for controllers designed using data collected over a limited operating range to maintain satisfactory control performance outside that range. 
We introduce a novel PI-D3C framework that integrates pure data-driven virtual reference feedback tuning (VRFT) with prior physical knowledge, thereby significantly improving extrapolation performance beyond the operating range covered by the collected data.
In the proposed approach, the gain-scheduled controller parameters are designed based on both training data and the yaw-rate response predicted by the physics-based model. To evaluate the effectiveness of the proposed method, numerical simulations are conducted using co-simulation between MATLAB/Simulink and CarSim. The results demonstrate that the controller designed by pure VRFT achieves satisfactory tracking performance within the training velocity region but exhibits significant deterioration in tracking performance outside the training region.
In contrast, the proposed controller maintains good yaw-rate tracking performance over a wider velocity range while suppressing excessive variations in the scheduled gains and steering input. These results indicate that incorporating prior physical knowledge into pure data-driven VRFT improves the extrapolation performance of gain-scheduled controllers designed using limited training data.
\end{abstract}


\section{Introduction}
\label{sec:introduction}
Vehicle lateral control systems are required to maintain satisfactory yaw-rate tracking performance under modeling uncertainties and variations in driving conditions~\cite{li_adaptive-event-trigger-based_2021,gao_robust_2022,chen_tube-based_2026,yahagi_data-driven_2025}. However, controller design for lateral vehicle dynamics is challenging because the vehicle dynamics exhibit nonlinear characteristics and vary depending on the driving conditions~\cite{zhang_vehicle_2018,zhu_real-time_2013}. In particular, changes in vehicle velocity significantly affect the tire lateral dynamic characteristics and yaw-rate response. Therefore, a fixed-gain controller may not achieve the desired control performance over a wide velocity range~\cite{kapsalis_gain-scheduled_2026,kapsalis_gainscheduled_2020,wang_gain_2018,hang_lpvh_2019}. In addition, acquiring training data that covers a wide range of driving conditions is difficult in practical vehicle systems. Steering tests that strongly excite the lateral acceleration and yaw rate may bring the vehicle close to its vehicle stability limits, and sufficiently rich excitation signals cannot always be applied because of safety constraints. 

In recent years, direct data-driven control methods, such as correlation-based tuning (CbT)~\cite{karimi_iterative_2004}, fictitious reference iterative tuning (FRIT)~\cite{soma_new_2004}, and virtual reference feedback tuning (VRFT)~\cite{campi_virtual_2002,campi_direct_2006}, have attracted increasing attention because they can tune controllers directly from measured input-output data without relying on a detailed plant model~\cite{karimi_iterative_2004,soma_new_2004,campi_virtual_2002,campi_direct_2006}. These approaches are useful for practical vehicle systems in which accurate mathematical modeling is difficult. In particular, VRFT can determine controller parameters through an optimization problem using a single set of input-output data. However, conventional VRFT-based methods strongly depend on the information contained in the training data. Therefore, when the training data are collected only under limited driving conditions, satisfactory control performance is not necessarily achieved in extrapolation regions that are not represented in the data. For vehicle yaw-rate control, this limitation is crucial because the system dynamics vary significantly with vehicle velocity. Previous studies have extended VRFT to gain-scheduled controllers~\cite{formentin_direct_2016,yahagi_oneshot_2026}, lookup-table-based controllers~\cite{yahagi_direct_2022-2}, gain-scheduled sparse polynomial controllers~\cite{yahagi_direct_2022}, and model-free controllers~\cite{rallo_vehicle_2016,radac_data-driven_2018,yahagi_non-iterative_2022}. These studies indicate that VRFT is effective for tuning controllers without identifying a plant model, even for systems whose characteristics vary depending on the operating condition. For vehicle yaw-rate control, a gain-scheduled controller is particularly effective because the controller gains can be varied according to the vehicle velocity~\cite{wang_gain_2018}. Nevertheless, pure data-driven gain-scheduled VRFT is constructed primarily from measured time-series data. Thus, when the training data does not sufficiently cover the target operating range, the obtained gain-scheduling law may suffer from poor extrapolation performance. 

Physics-informed machine learning (PIML) has recently attracted considerable attention as a framework for integrating measured data with mathematical models and physical laws~\cite{karniadakis_physics-informed_2021}. In physics-informed approaches, prior physical knowledge is incorporated into the learning process to supplement limited or noisy data, thereby improving the accuracy and generalization capability of the resulting models. 
Although PIML has primarily been studied in the context of predictive model identification using neural networks~\cite{de_giuli_physics-informed_2024,raissi_physics-informed_2019} and kernel-based regression methods~\cite{doumeche_physics-informed_nodate,thorpe_physics-informed_2024}, its fundamental concept of combining data with available physical knowledge can also provide a useful basis for data-driven controller design~\cite{koyanagi_data-driven_2026}. In this context, incorporating an approximate physical model into the controller-tuning criterion is expected to compensate for information that is insufficiently represented in the measured data. 
Unlike conventional PIML approaches, the framework considered in this paper does not train a surrogate model subject to physical constraints. Instead, an approximate physics-based vehicle model is embedded in the controller-tuning objective as an auxiliary means of evaluating closed-loop performance. Although the fundamental concept was presented in our preliminary conference paper~\cite{koyanagi_data-driven_2026}, that study was limited to the design of a fixed-gain PID controller. 

This study proposes a physics-informed direct data-driven controller design method (PI-D3C) for gain-scheduled proportional-integral (PI) control of vehicle yaw-rate dynamics. In the proposed method, prior physical knowledge derived from a single-track bicycle model is incorporated into the pure data-driven VRFT cost function. The conventional VRFT criterion is constructed exclusively from measured input-output data, whereas the additional physics-informed terms evaluate the closed-loop reference model tracking performance predicted by the bicycle model. The controller gains are scheduled according to vehicle velocity, and the gain-scheduling parameters are determined by simultaneously considering the measured data and the physics-based model responses. Consequently, the proposed method is expected to retain satisfactory performance within the training velocity region while improving reference model tracking performance in velocity regions that are insufficiently represented in the training data. The effectiveness of the proposed method is verified through numerical simulations of lateral vehicle dynamics using CarSim. The proposed physics-informed VRFT-based gain-scheduled PI controller is compared with a pure data-driven VRFT-based gain-scheduled PI controller. The comparison is conducted in terms of yaw-rate tracking performance, steering input, gain-scheduling characteristics, and extrapolation performance under vehicle-velocity variations.  

The remainder of this paper is organized as follows. Section~\ref{sec:preliminaries} presents the preliminaries, including the notation, problem setting, and gain-scheduled PI control. Section~\ref{sec:proposed_method} describes the proposed physics-informed VRFT-based gain-scheduled PI control method. Section~\ref{sec:simulation} presents numerical simulations to verify the effectiveness of the proposed method and discusses the results. Section~\ref{sec:conclusion} concludes this paper.

\section{Preliminaries}
\label{sec:preliminaries}

\subsection{Notation}
Let \(\mathbb{R}\) denote the set of real numbers. The set of
real-valued matrices of dimension \(m \times n\) is denoted by
\(\mathbb{R}^{m \times n}\), and \(\mathbb{R}^{m}\) denotes
\(\mathbb{R}^{m \times 1}\). For a vector \(x\),
\(\left\| x \right\|_{2}\) denotes its Euclidean norm. The identity
matrix is denoted by \(\mathcal{I}\). The signal spaces of the reference
input, plant input, plant output, tracking error, and scheduling
variable are denoted by \(\mathcal{R}\), \(\mathcal{U}\),
\(\mathcal{Y}\), \(\mathcal{E}\), and \(\mathcal{V}\), respectively. A
discrete-time signal \(x\) over a finite interval is represented by
\(x = \left\{ x(t) \right\}_{t = 1}^{N}.\)

\subsection{Problem Formulation}
\begin{figure}[!t]
	\centering
	\includegraphics[width=0.75\columnwidth]{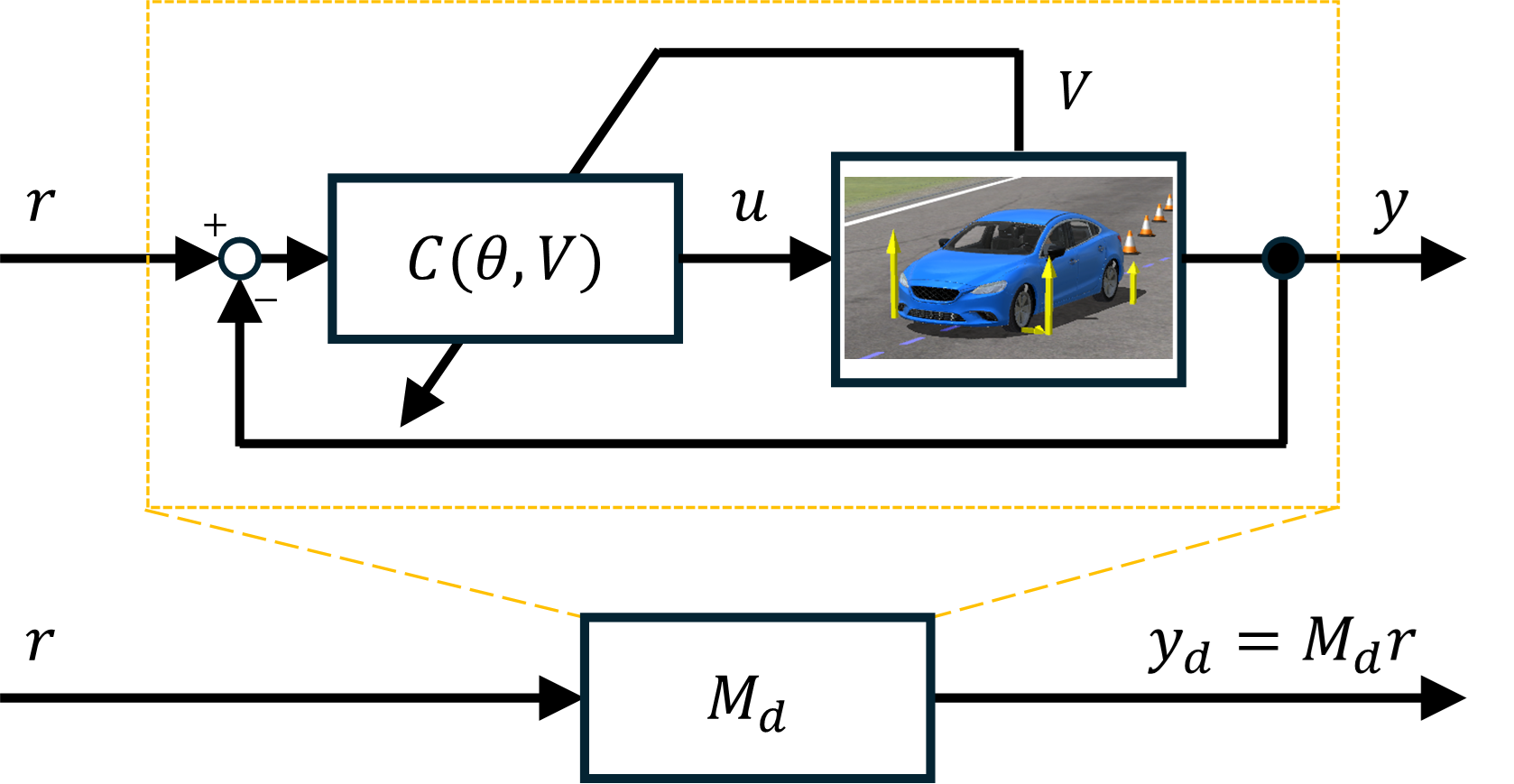}
	\caption{Model-reference vehicle lateral gain-scheduled control.}
	\label{fig:MRC_GS_control}
\end{figure}

We consider a model-reference gain-scheduled control problem for the
nonlinear lateral dynamics of a vehicle, as illustrated in Fig.~\ref{fig:MRC_GS_control}. 
The steering angle \(u\in\mathbb{R}\) is the control input, the yaw rate
\(y\in\mathbb{R}\) is the controlled output, the target yaw rate
\(r\in\mathbb{R}\) is the reference input, and the vehicle speed
\(v\in\mathbb{R}\) is used as the scheduling variable. The nonlinear
vehicle dynamics are represented by a causal input-output operator
\(P:\mathcal{U}\rightarrow\mathcal{Y}\), such that \(y = P(u)\). The effects of
the initial state and disturbances are omitted from the notation for
simplicity. The tracking error is defined as \(e = r - y\). The
gain-scheduled feedback controller is represented by a causal operator
\(C_{\theta}:\mathcal{E}\times\mathcal{V}\rightarrow\mathcal{U},\) where
\(\theta \in \Theta \subseteq \mathbb{R}^{n_{\theta}}\) is the tunable
controller parameter vector. The control input is given by
\(u = C_{\theta}(e,v)\). The parameter vector \(\theta\) remains fixed
during controller operation and determines the scheduling law, whereas
the actual controller gains may vary with the scheduling variable
\(v(t)\). The operator representation also includes dynamic
gain-scheduled controllers whose outputs depend on the histories of the
tracking error and scheduling variable. The closed-loop interconnection
of the plant and controller is described by
\(y = P\left( C_{\theta}\left( r - y,v \right) \right)\). Its value at
time \(t\) is denoted by \(y\left( t;\theta \right)\). The
designer-specified reference model is represented by the causal operator
\(M_{d}\mathcal{:R \rightarrow Y}\). For a given reference input \(r\),
the desired response is defined as \(y_{d} = M_{d}(r)\) and its value at
time \(t\) is denoted by \(y_{d}(t)\).

The objective of the model-reference control problem is to determine
\(\theta\) such that the closed-loop input-output behavior of the
nonlinear vehicle coincides with that specified by \(M_{d}\). For a
finite evaluation interval \(t = 1,\ldots,N\), the model-reference cost
function is defined as
\begin{equation} 
	J_{MR}(\theta) = \frac{1}{N}\sum_{t = 1}^{N} \left\lVert y(t;\theta)-y_d(t) \right\rVert_2^2. 
\end{equation}

Direct minimization of this cost function generally requires either an
explicit mathematical model of the nonlinear vehicle dynamics or
repeated closed-loop experiments for different candidate values of
\(\theta\). In this study, we adopt the data-driven approach without
explicitly identifying the plant operator \(P\). When the measured data
do not adequately cover the target operating range, a controller tuned
solely from those data may exhibit degraded performance outside the
training region. In this study, we achieve improved control performance
outside the learning region by incorporating prior physical knowledge
into data-driven control.

\subsection{Gain-Scheduled PI Controller}
\label{subsec:GS_PI_Ctrl}
The gain-scheduled PI controller \(C_{\theta_{PI}}\) is defined in
discrete time by

\begin{equation}
	u(t) = K_{P}\left( v(t) \right)e(t) + K_{I}\left( v(t) \right)\xi(t)
\end{equation}
\begin{equation}
	\xi(t + 1) = \xi(t) + T_{s}e(t)
\end{equation}
where \(e(t)\) denotes the tracking error, \(\xi(t)\) denotes the
integral state, and \(T_{s}\) is the sampling period. \(v(t)\) is
normalized vehicle velocity, given as

\[v(t) = \frac{V(t)}{V_{n}}\]
where \(V(t)\) is the physical vehicle speed and \(V_{n}\) is the
normalization speed. The proportional and integral gains are
parameterized as polynomial functions of the vehicle velocity:

\[K_{P}\left( v(t) \right) = w_{P}^{\top}x_{f}\left( v(t) \right)\]
\[K_{I}\left( v(t) \right) = w_{I}^{\top}x_{f}\left( v(t) \right)\]
where \(x_{f}(v)\) is the basis-function vector of the scheduling
variable. In this study, a third-order polynomial basis is used:
\[x_{f}(v) = \begin{bmatrix}
1 & v & v^{2} & v^{3}
\end{bmatrix}^{\top}.\]
The coefficient vectors that determine the proportional and integral
gain-scheduling laws are defined as
\[w_{P} = \begin{bmatrix}
w_{P0} & w_{P1} & w_{P2} & w_{P3}
\end{bmatrix}^{\top}\]
\[w_{I} = \begin{bmatrix}
w_{I0} & w_{I1} & w_{I2} & w_{I3}
\end{bmatrix}^{\top}.\]
From the above, the control input is given as
\[
u(t) = w_{P}^{\top}x_{f}\left( v(t) \right)e(t) + w_{I}^{\top}x_{f}\left( v(t) \right)\xi(t).
\]
By defining
\[\theta_{PI} = \begin{bmatrix}
w_{P} \\
w_{I}
\end{bmatrix} \in \mathbb{R}^{8}\]
and
\[
\varphi(t) = \begin{bmatrix}
x_{f}\left( v(t) \right)e(t) \\
x_{f}\left( v(t) \right)\xi(t)
\end{bmatrix} \in \mathbb{R}^{8},
\]
the control input is given as
\begin{equation}
	u(t) = \varphi^{\top}(t)\theta_{PI}.
\end{equation}
The parameter vector \(\theta_{PI}\) is fixed during controller
operation and determines the gain-scheduling laws. In contrast, the
actual gains \(K_{P}\left( v(t) \right)\) and
\(K_{I}\left( v(t) \right)\) vary with the vehicle velocity.

\subsection{Nominal Model of Vehicle Dynamics}
\label{subsec:nominal_model}

\begin{figure}[!t]
\centering
\includegraphics[width=0.6\columnwidth]{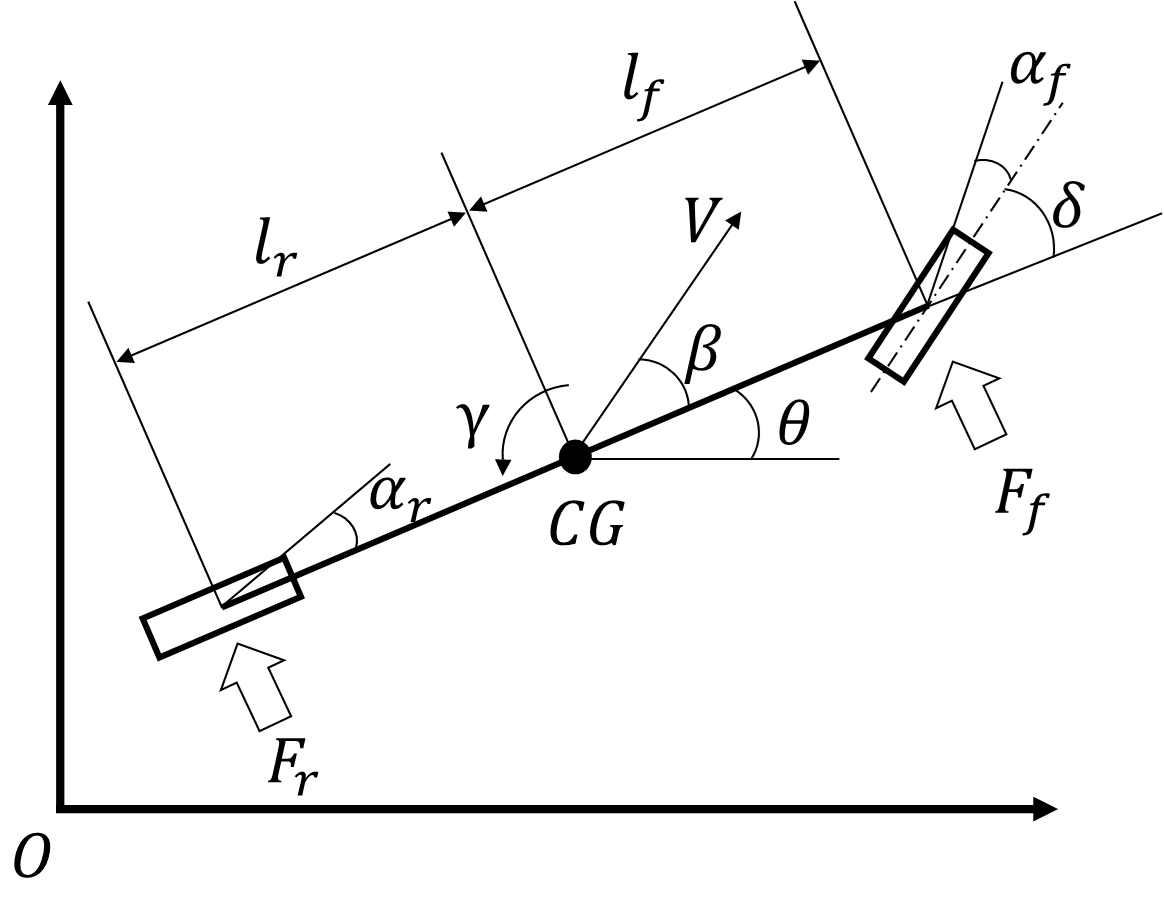}
\caption{Single-track bicycle model.}
\label{fig:Bicycle_model}
\end{figure}

In the proposed approach, the bicycle model is used as partial physical knowledge of the actual vehicle dynamics to be controlled. Fig.~\ref{fig:Bicycle_model} shows the single-track bicycle model. The state vector of the bicycle model is defined as

\[
x_{p}(t) = \begin{bmatrix}
\gamma(t) \\
\beta(t)
\end{bmatrix}
\]
where \(\gamma\) and \(\beta\) denote the yaw rate and vehicle sideslip
angle, respectively. The continuous-time vehicle dynamics are
represented by
\begin{equation} 
	P_n: \left\{ 
		\begin{aligned} 
			\dot{x}_p(t) &= A\!\left(V(t)\right)x_p(t) + B\!\left(V(t)\right)u(t) \\
			\hat{y}(t) &= Cx_p(t)
		\end{aligned} 
		\right. 
\end{equation}
where \(u(t)\) is the front steering angle, \(y(t)\) is the yaw rate,
and \(V(t)\) is the vehicle velocity. The hat notation distinguishes the
output predicted by the nominal physical model from the output
\(y(t)\) of the controlled plant. The velocity-dependent system matrix
is given by
\begin{equation}
A(V) = \begin{bmatrix}
 - \frac{2\left( l_{f}^{2}C_{f} + l_{r}^{2}C_{r} \right)}{I_{z}V} & - \frac{2\left( l_{f}C_{f} - l_{r}C_{r} \right)}{I_{z}} \\
 - 1 - \frac{2\left( l_{f}C_{f} - l_{r}C_{r} \right)}{MV^{2}} & - \frac{2\left( C_{f} + C_{r} \right)}{MV}
\end{bmatrix}
\end{equation}
%
%
and the input and output matrices are given by
\begin{equation}
B(V) = \begin{bmatrix}
\frac{2l_{f}C_{f}}{I_{z}} \\
\frac{2C_{f}}{MV}
\end{bmatrix},\ C = \begin{bmatrix}
1 & 0
\end{bmatrix}
\end{equation}
where \(M\) is the vehicle mass, \(I_{z}\) is the yaw moment of inertia,
\(l_{f}\) and \(l_{r}\) are the distances from the vehicle center of
gravity to the front and rear axles, respectively, and \(C_{f}\) and
\(C_{r}\) are the cornering powers of the front and rear tires,
respectively. During the closed-loop simulation, the system matrices of
the bicycle model are updated according to the instantaneous vehicle
speed \(V(t)\).

\section{Physics-Informed Direct Data-Driven Gain-Scheduled PI Control}

\label{sec:proposed_method}

This section presents the proposed physics-informed direct data-driven tuning
method for the gain-scheduled PI controller introduced in Section~\ref{subsec:GS_PI_Ctrl}.
We first present an overview of the proposed method and then describe
its data-driven and physics-based components.

\subsection{Overview}
\begin{figure}[!t]
\centering
\includegraphics[width=0.95\columnwidth]{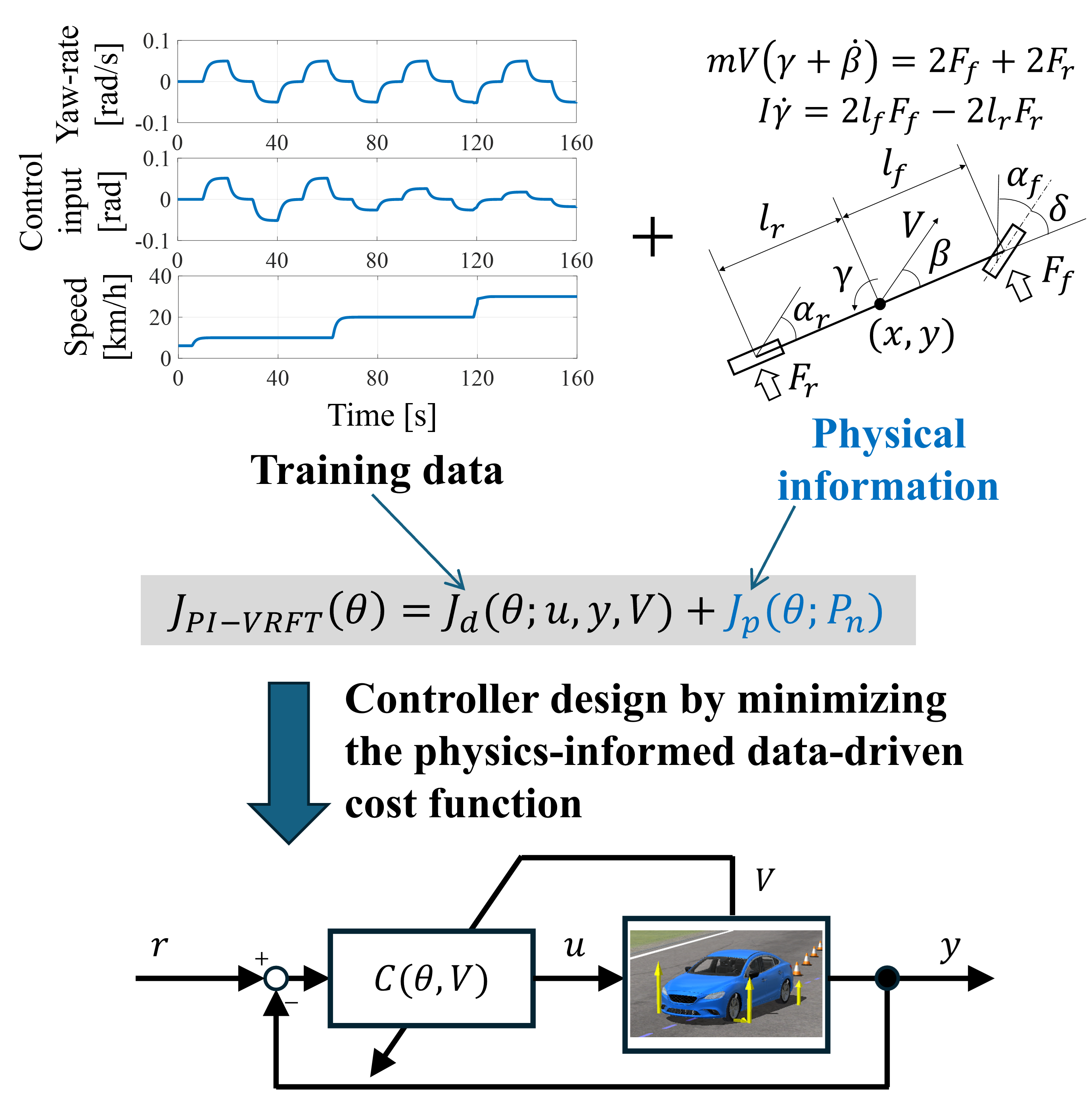}
\caption{
	Overview of the proposed physics-informed direct data-driven control (PI-D3C) framework, specifically the PI-VRFT framework.
	}
\label{fig:Overview_proposed_method}
\end{figure}

The overview of the proposed method is shown in Fig.~\ref{fig:Overview_proposed_method}. The proposed cost function of the 
physics-informed VRFT (i.e., PI-D3C) is defined as

\begin{equation}
	\label{eq:physcs-infromed cost function}
J_{PI\text{-}VRFT}\left( \theta_{PI} \right) = J_{d}\left( \theta_{PI} \right) + J_{p}\left( \theta_{PI} \right)
\end{equation}
where \(J_{d}\) is the data-driven VRFT criterion constructed from the
measured input-output data, and \(J_{p}\) is the physics-based criterion
evaluated using the auxiliary bicycle model. The physics-based criterion
includes separate weighting coefficients for the interpolation and
extrapolation regions, as described in Section~\ref{subsec:physics_based_cost_function}.

Let the measured data be given by
\(
\mathcal{D}_{N} = \left\{ u_{0}(t), y_{0}(t), v_{0}(t) \mid t = 1, \ldots, N_{train} \right\}
\)
where \(u_{0}\), \(y_{0}\), and \(v_{0}\) denote the measured control
input, plant output, and scheduling signal, respectively. For safety
reasons, the speed range in which testing with real vehicles can be
conducted is limited. Consequently, the available training data cover only a limited speed range.
The lower and upper bounds of the training velocity range are defined as
\[
v^{\min}_{train} = \min_{v \in \mathcal{V}_{train}}v
\]
\[
v^{\max}_{train} = \max_{v \in \mathcal{V}_{train}}v.
\]
Accordingly, the training velocity region is defined as
\[
\Omega_{train} = \left\lbrack v^{\min}_{train},v^{\max}_{train} \right\rbrack.
\]
To evaluate candidate controllers beyond the operating conditions
represented in the training data, a physical vehicle-speed profile
\[
V_{eval}(t),\ t = 1,\ldots,N_{p}
\]
is introduced, where \(N_{p}\) denotes the number of time samples used
in the physics-based closed-loop simulation. The corresponding
normalized scheduling profile is given by
\[
v_{eval}(t) = \frac{V_{eval}(t)}{V_{n}}.
\]
The evaluation profile may pass through both the training speed region
and speed regions outside it. The instantaneous tracking errors are
therefore classified into interpolation and extrapolation components
according to \(v_{eval}(t)\).

\subsection{Data-driven cost function}
The data-driven cost function is introduced by applying VRFT which is
one of data-driven control methods to achieve the model-matching
criteria (1) from dataset \(\mathcal{D}_{N}\). 
According to~\cite{yahagi_direct_2022}, the gain-scheduled data-driven cost function is derived.
In VRFT, the virtual reference signal is introduced as
\(r_{v} = M_{d}^{- 1}\left( y_{0} \right)\). The corresponding virtual
tracking error is given by
\(\widetilde{e} = r_{v} - y_{0} = \left( M_{d}^{- 1}\mathcal{- I} \right)y_{0}\).
Accordingly, the VRFT cost function is defined as
\begin{equation}
J_{VR}(\theta) = \frac{1}{N}\sum_{t = 1}^{N} \left\lVert Lu_{0}(t) - LC_{\theta}( \widetilde{e},v_{0}) \right\lVert_{2}^{2}
\end{equation}
where \(L\) is the prefilter whose role is to make the VRFT cost
function equivalent, or as close as possible, to the model-reference
criterion. This cost function is constructed from the measured data and
the desired reference model. Specifically, the data-driven VRFT
criterion for the gain-scheduled PI controller is defined as

\begin{equation}
 J_{d}\left( \theta_{PI} \right) = \frac{1}{N}\sum_{t = 1}^{N} \left\lVert Lu_{0}(t) - L\varphi_{VRFT}^{\top}(t)\theta_{PI} \right\lVert_{2}^{2}
\end{equation}
with
\[
\varphi_{VRFT}(t) = \begin{bmatrix}
x_{f}\left( v_{0}(t) \right)\widetilde{e}(t) \\
x_{f}\left( v_{0}(t) \right)\widetilde{\xi}(t)
\end{bmatrix} \in \mathbb{R}^{8}
\]
\[
\widetilde{\xi}(t + 1) = \widetilde{\xi}(t) + T_{s}\widetilde{e}(t),\ \widetilde{\xi}(1) = 0.
\]
\rem{In pure data-driven tuning such as VRFT, the
controller parameters are determined only from measured input-output
data. Therefore, when the training data are collected under limited
operating conditions, the resulting gain-scheduled controller is
directly constrained only within the scheduling-variable range covered
by the training data. Consequently, its closed-loop performance may
deteriorate outside this range. To address this challenge, the proposed
method evaluates candidate controller parameters using a bicycle model
over both the interpolation and extrapolation regions and incorporates
the predicted model-reference tracking performance into the
controller-tuning criterion.}

\subsection{Physics-based cost function}
\label{subsec:physics_based_cost_function}
To evaluate a candidate controller under operating conditions that are
insufficiently represented in the measured training data, a bicycle
model, as shown in Section~\ref{subsec:nominal_model}, is introduced as auxiliary prior
physical knowledge. The bicycle model is not intended to reproduce the
controlled plant exactly. Instead, it is used to evaluate whether a
candidate gain-scheduled PI controller provides reasonable closed-loop
tracking performance over the vehicle-speed range of interest.

For the physics-based evaluation, the reference input profile and the
vehicle-speed profile are defined as
\[
r_{eval} = \left\{ r_{eval}(t) \right\}_{t = 1}^{N_{p}}
\]
\[
V_{eval} = \left\{ V_{eval}(t) \right\}_{t = 1}^{N_{p}}
\]
where \(N_{p}\ \)denotes the number of samples used in the physics-based
closed-loop simulation. The desired output corresponding to the
evaluation reference profile is defined as
\[
y_{d,eval} = M_{d}\left( r_{eval} \right)
\]
and its value at time \(t\) is denoted by \(y_{d,eval}(t)\).

For a candidate controller parameter vector \(\theta_{PI}\), the
reference input profile \(r_{eval}\) and the vehicle-speed profile
\(V_{eval}\) are applied to the closed-loop interconnection of the
gain-scheduled PI controller \(C_{\theta_{PI}}\) and the bicycle
model\(\ P_{n}\). The closed-loop response is given as
\(\hat{y} = P_{n}\lbrack C_{\theta_{PI}}\left( r_{eval} - \hat{y},v_{eval} \right)\rbrack\).
From the above, the resulting output predicted by the bicycle model is
denoted by \(\hat{y}\left( t;\theta_{PI} \right).\)

The evaluation vehicle-speed profile may pass through both the training
speed region and speed regions outside it. To distinguish between the
interpolation and extrapolation regions along the evaluation profile,
the following indicator functions are introduced:

\[\mathbb{I}_{in}(t) = \left\{ \begin{matrix}
1, & V^{\min}_{train} \leq V_{eval}(t) \leq V^{\max}_{train} \\
0, & \text{otherwise}
\end{matrix} \right.\ \]
and
\[
\mathbb{I}_{out}(t) = 1 - \mathbb{I}_{in}(t)
\]
where \(V^{\min}_{train}\) and \(V^{\max}_{train}\) denote the lower and
upper bounds of the vehicle-speed range represented in the measured
training data, respectively.

The numbers of samples belonging to the interpolation and extrapolation
regions are defined as
\[
N_{in} = \sum_{t = 1}^{N_{p}}\mathbb{I}_{in}(t),\ N_{out} = \sum_{t = 1}^{N_{p}}\mathbb{I}_{out}(t),
\]
respectively. Provided that \(N_{in} > 0\), the physics-based tracking
criterion over the interpolation region is defined as
\[
J_{p,in}\left( \theta_{PI} \right) = \frac{1}{N_{in}}\sum_{t = 1}^{N_{p}}\mathbb{I}_{in}(t)\left\lVert y_{d,eval}(t) - \hat{y}\left( t;\theta_{PI} \right) \right\lVert_{2}^{2}.
\]
Similarly, provided that \(N_{out} > 0\), the physics-based tracking
criterion over the extrapolation region is defined as
\[
J_{p,out}\left( \theta_{PI} \right) = \frac{1}{N_{out}}\sum_{t = 1}^{N_{p}}\mathbb{I}_{out}(t)\left\lVert y_{d,eval}(t) - \hat{y}\left( t;\theta_{PI} \right) \right\lVert_{2}^{2}.
\]
The overall physics-based criterion is defined as
\[
J_{p}\left( \theta_{PI} \right) = \lambda_{in}J_{p,in}\left( \theta_{PI} \right) + \lambda_{out}J_{p,out}\left( \theta_{PI} \right)
\]
where \(\lambda_{in} \geq 0\) and \(\lambda_{out} \geq 0\) are weighting
coefficients for the interpolation and extrapolation regions,
respectively. A larger value of \(\lambda_{out}\) places greater
emphasis on the predicted closed-loop tracking performance outside the
vehicle-speed range represented in the measured training data. If the
evaluation profile contains samples from only one of the two regions,
only the criterion corresponding to the nonempty region is included in
\(J_{p}\). From the above, the controller parameter vector is determined
by solving
\[
\theta_{PI}^{\star} = \arg{\min_{\theta_{PI} \in \Theta}{J_{PI\text{-}VRFT}\left( \theta_{PI} \right)}}.
\]
\rem{Unlike the conventional VRFT criterion for the
gain-scheduled PI controller, the physics-based criterion
\(J_{p}\left( \theta_{PI} \right)\) is generally nonlinear with respect
to \(\theta_{PI}\). This is because the predicted output
\(\hat{y}\left( t;\theta_{PI} \right)\) is obtained through a
closed-loop simulation of the bicycle model. Therefore, the optimal
controller parameter vector is determined using a numerical optimization
method.}

\rem{The physical model serves as an auxiliary regularizer
that discourages candidate gain-scheduling laws from producing
undesirable closed-loop responses in vehicle-speed regions that are
poorly represented or absent in the measured training data.}

\section{Simulation}
\label{sec:simulation}

\subsection{Simulation Conditions}
\begin{figure}[!t]
\centering
\includegraphics[width=0.4\columnwidth]{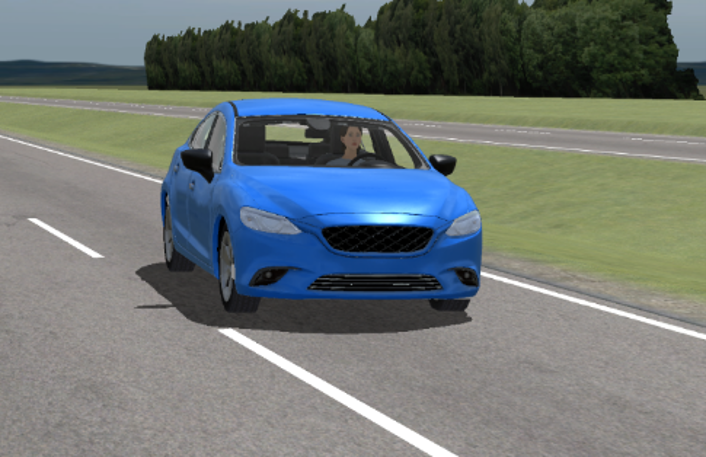}
\caption{The vehicle used in the simulation tests.}
\label{fig:Target_Vehicle}
\end{figure}

This subsection describes the simulation conditions employed to evaluate
the effectiveness of the proposed physics-informed VRFT-based
gain-scheduled PI controller for vehicle yaw-rate tracking control. Fig.~\ref{fig:Target_Vehicle} shows the controlled plant, which is a D-Class Sedan vehicle model implemented in CarSim. 
The specifictions of the nominal vehicel model (i.e., bicycle model) are
listed in Table~\ref{tab:vehicle_specifications}. As
described in Section~\ref{subsec:nominal_model}, the cornering power values are given for the linear tire region. 
The control input is the front steering angle, and
the controlled output is the yaw rate. The vehicle velocity is used as
the scheduling variable of the gain-scheduled PI controllers.

\begin{table}[t] 
	\centering 
	\caption{Nominal vehicle model specifications} 
	\label{tab:vehicle_specifications} 
	\begin{tabular}{clcc} 
		\toprule
		Symbol & Description & Unit & Value \\
		\midrule 
		$M$ & Vehicle mass & kg & 1613 \\ 
		$I_z$ & Vehicle inertia moment & kg\,m$^2$ & 2967 \\ 
		$l_f$ & Length from CG to front axle & m & 1.137 \\ 
		$l_r$ & Length from CG to rear axle & m & 1.729 \\ 
		$C_f$ & Cornering power of front tire & N/rad & $7.100 \times 10^4$ \\ 
		$C_r$ & Cornering power of rear tire & N/rad & $4.850 \times 10^4$ \\
		\bottomrule
	\end{tabular} 
\end{table}

The input-output training data used for controller tuning are obtained
while the vehicle velocity is varied up to 30~km/h. A
fixed-gain PI controller with initial gains of \(K_{P} = 0.3\) and
\(K_{I} = 0.2\) is used to collect the training data. The same training
dataset is used to tune both the conventional VRFT-based gain-scheduled
PI controller and the proposed physics-informed VRFT-based
gain-scheduled PI controller.

After controller tuning, numerical simulations are conducted while the
vehicle velocity is varied up to 80~km/h. Therefore, the
velocity range from 30~km/h to 80~km/h is not included in the training
data and is regarded as an extrapolation region. This condition is
introduced to evaluate whether the proposed method can improve the
control performance outside the velocity range represented in the
training data.

The desired reference model is defined as the following second-order
system:
\[
M_{d}\mathcal{= Z}\left\lbrack \left( \frac{1}{0.5s + 1} \right)^{2} \right\rbrack
\]
where \(\mathcal{Z} \left\lbrack\cdot \right\rbrack\) denotes the discretization operator.
The simulation sampling period is set to 0.001~s, whereas the controller sampling period is set to 0.005~s. The parameters $\lambda_{\mathrm{in}}$ and $\lambda_{\mathrm{out}}$ are both set to 1. 
The cost function \eqref{eq:physcs-infromed cost function} is minimized using the MATLAB \texttt{fmincon} function.

\begin{figure}[!t]
\centering
\includegraphics[width=0.83\columnwidth]{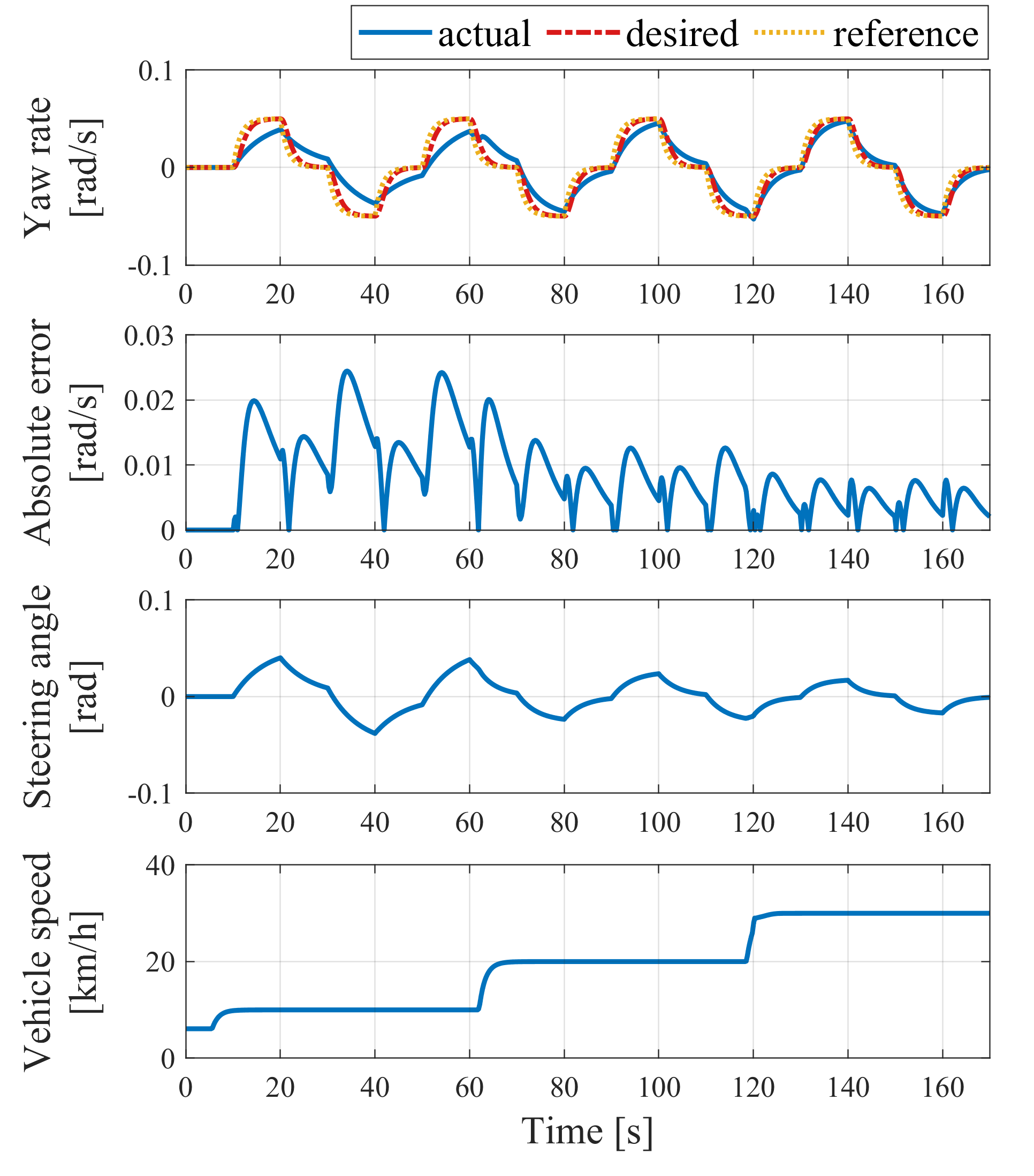}
\caption{Training data.}
\label{fig:Training_data}
\end{figure}

Fig.~\ref{fig:Training_data} shows the input-output training data used for VRFT-based controller tuning. 
The upper plot shows the target yaw rate, the desired reference model response, and the output yaw rate. The middle plot shows the control input, namely the front steering angle, and the lower plot shows the vehicle velocity. 
The parameters of the conventional gain-scheduled PI controller are determined by minimizing the conventional VRFT cost function. In contrast, the parameters of the
proposed controller are determined by minimizing the physics-informed
VRFT cost function, which consists of the conventional VRFT term and the
physics-based terms derived from the bicycle model. The same controller
structure, training data, reference model, and gain-scheduling basis
functions are used for both methods to ensure a fair comparison. The
control performance is evaluated based on the tracking error between the
output yaw rate and the desired reference model response. The mean
squared error (MSE) of the tracking error is defined as
\(J_{MSE} = \frac{1}{N_{v}}\sum_{k = 1}^{N_{v}}\left\| y(k) - y_{d}(k) \right\|_{2}^{2}\),
where \(N_{v}\) is the number of samples in the validation simulation.
The tracking performance of the conventional and proposed methods is
compared over the entire validation range, with particular attention
paid to the extrapolation region above 30~km/h.

\subsection{Results and Discussion}

\begin{figure}[!t]
\centering
\includegraphics[width=0.83\columnwidth]{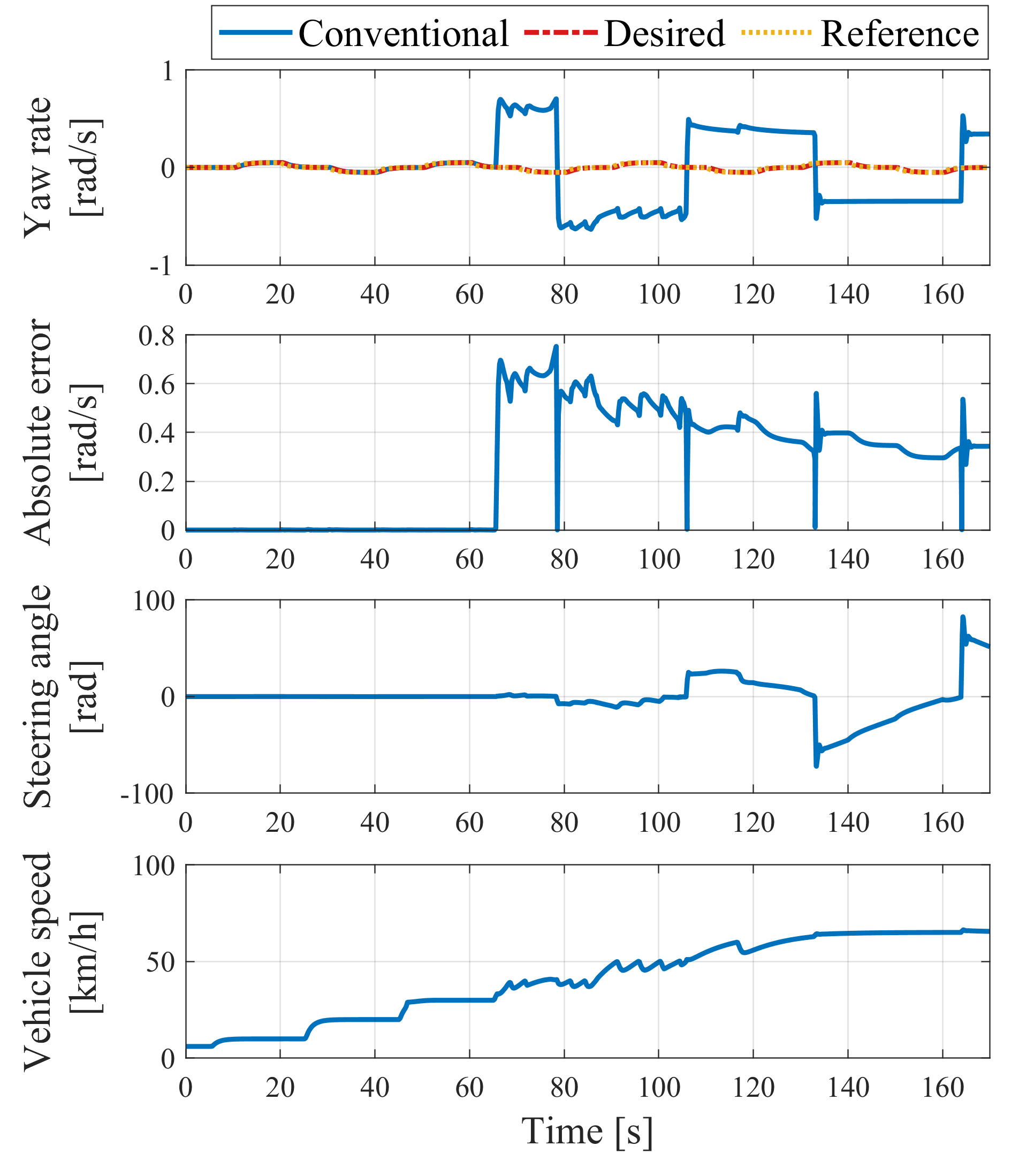}
\caption{Simulation results of the standard VRFT-based gain-scheduled PI controller.}
\label{fig:SiimResult_pure_GS_VRFT}
\end{figure}

\begin{figure}[!t]
\centering
\includegraphics[width=0.83\columnwidth]{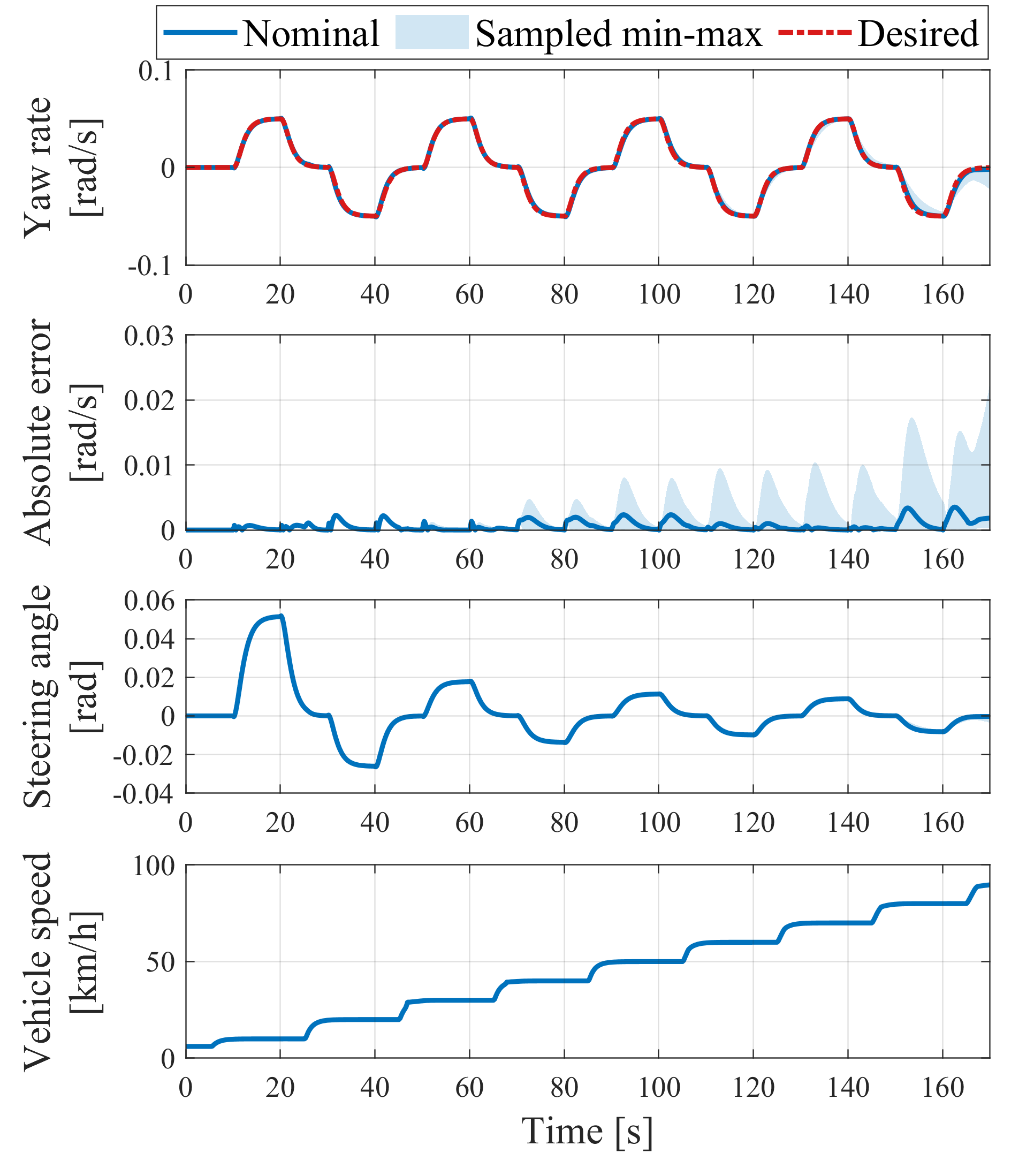}
\caption{
	Simulation results of the proposed physics-informed VRFT-based gain-scheduled PI controller under $\pm20\%$ parameter variations. The shaded region represents the sampled minimum and maximum responses from 100 randomly generated models.
}
\label{fig:SiimResult_proposed_GS_VRFT}
\end{figure}

Fig.~\ref{fig:SiimResult_pure_GS_VRFT} and Fig.~\ref{fig:SiimResult_proposed_GS_VRFT} show the simulation results obtained using the conventional VRFT-based gain-scheduled PI controller and the proposed physics-informed VRFT-based gain-scheduled PI controller, respectively. 
\textcolor{black}{
	In Fig.~\ref{fig:SiimResult_proposed_GS_VRFT}, the results are presented for both the nominal model specified in Table~\ref{tab:vehicle_specifications} and models with parameter variations of up to $\pm20\%$ from their nominal values. To evaluate robustness against parameter uncertainty, 100 models were generated by randomly perturbing the nominal model parameters within this range. The shaded regions defined by the minimum and maximum responses of the sampled models are also shown.
} 
In each figure, the upper plot shows the target yaw rate, the reference model response, and the vehicle yaw rate. The middle plot shows the control input, namely the front steering angle, and the lower plot shows the vehicle velocity.
As shown in Fig.~\ref{fig:SiimResult_pure_GS_VRFT}, the conventional gain-scheduled PI controller provides satisfactory tracking performance
within the training velocity region up to 30~km/h. 
In this region,
the output yaw rate closely follows the desired reference model
response. However, when the vehicle velocity exceeds the training region
and increases from 30 to 80~km/h, an excessively large steering input is
generated. Consequently, the yaw-rate response deviates significantly
from the desired reference model response, and satisfactory tracking
performance cannot be maintained in the untrained velocity region. In
contrast, Fig.~\ref{fig:SiimResult_proposed_GS_VRFT} shows that the proposed physics-informed gain-scheduled PI controller achieves satisfactory yaw-rate tracking performance within the training velocity region. 
Furthermore, the proposed controller maintains good tracking performance even in the untrained velocity region from 30 to 80 km/h. 
Although the controller parameters are identified using limited input-output data, the incorporation of the physics-based vehicle model into the cost function suppresses the deterioration in control performance outside the training velocity region. 
While the absolute error remains very small in the data-driven region (up to 30~km/h), it increases in the extrapolation region owing to modeling errors.
The MSE of the conventional gain-scheduled PI controller is
\(1.272 \times 10^{- 1}\), whereas that of the proposed physics-informed
gain-scheduled PI controller using the nominal parameters is \(9.913 \times 10^{- 7}\). 
Thus, the proposed method reduced the MSE by more than five orders of magnitude
compared with the conventional method. This result indicates that
incorporating prior physical knowledge into VRFT significantly improves
reference model tracking performance over the entire velocity range,
particularly in the untrained region.

\begin{figure}[!t]
\centering
\includegraphics[width=0.83\columnwidth]{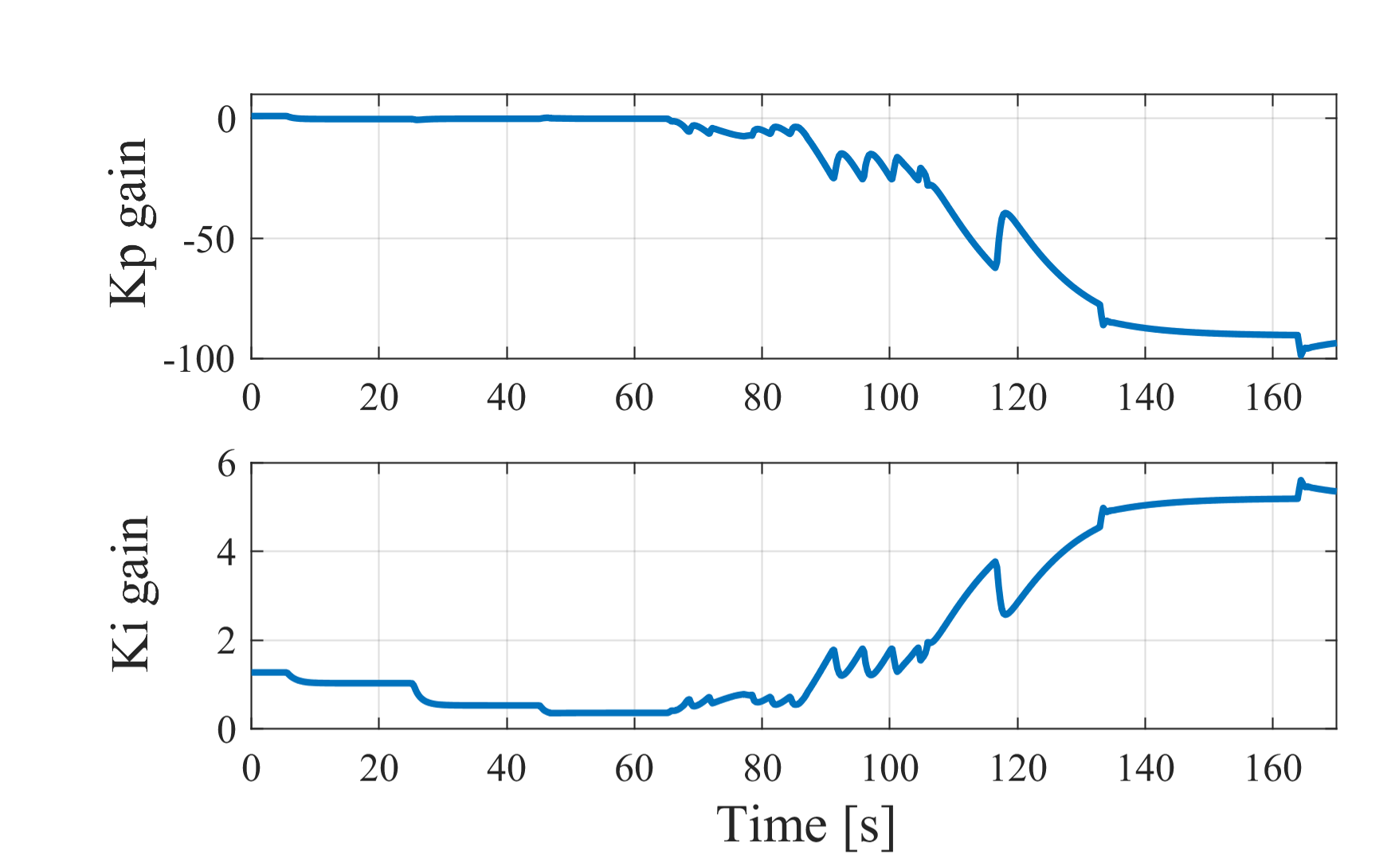}
\caption{Gain-scheduling characteristics of the VRFT-based gain-scheduled PI controller.}
\label{fig:SiimResult_pure_GS_VRFT_gain_characteristics}

\centering
\includegraphics[width=0.83\columnwidth]{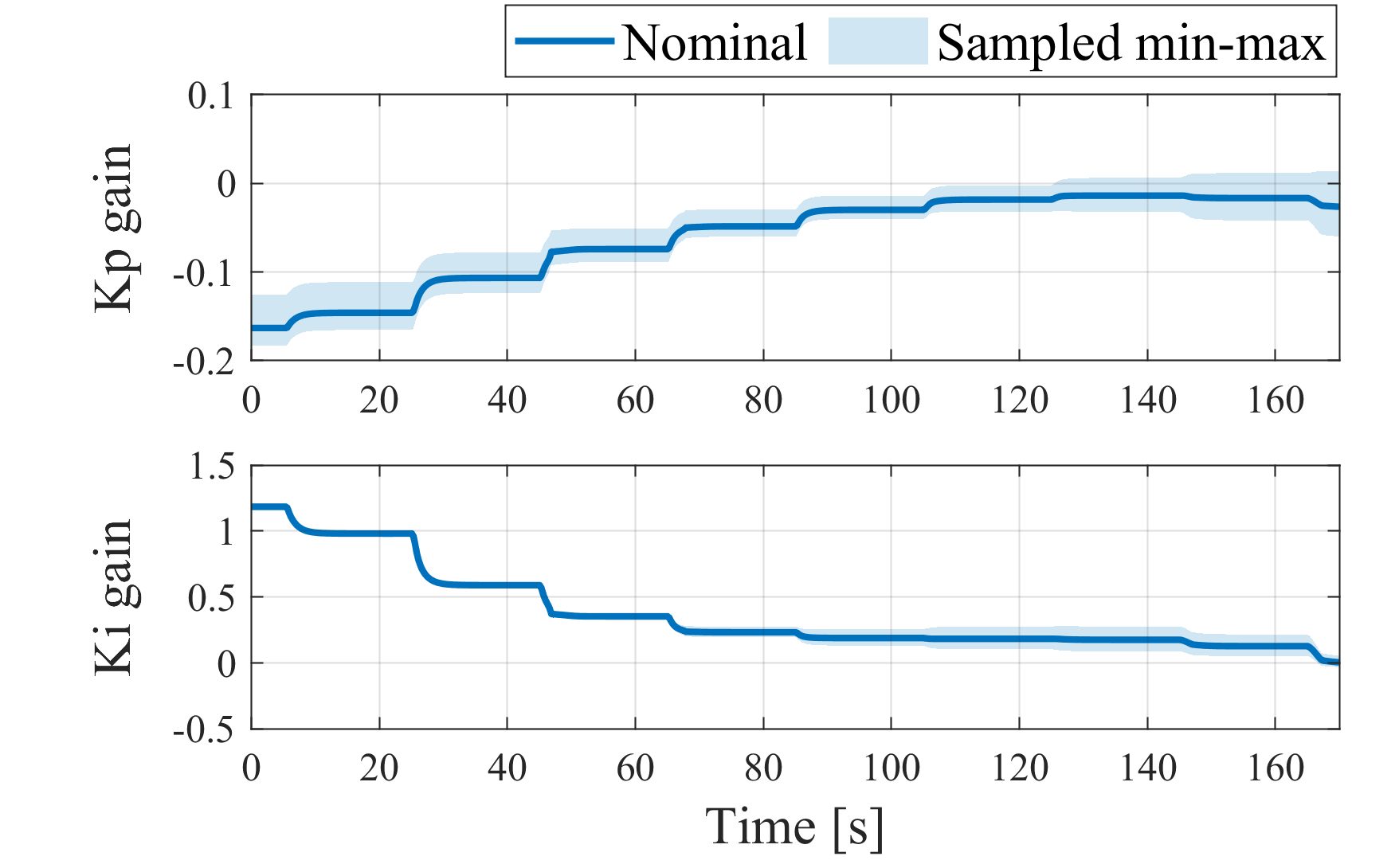}
\caption{
	Gain schedules of the proposed physics-informed VRFT-based gain-scheduled PI controller under $\pm20\%$ parameter variations.
}
\label{fig:SiimResult_proposed_GS_VRFT_gain_characteristics}
\end{figure}

Fig.~\ref{fig:SiimResult_pure_GS_VRFT_gain_characteristics} and Fig.~\ref{fig:SiimResult_proposed_GS_VRFT_gain_characteristics} show the gain-scheduling characteristics of the conventional and proposed gain-scheduled PI controllers, respectively.
\textcolor{black}{
	Fig.~\ref{fig:SiimResult_proposed_GS_VRFT_gain_characteristics} shows the scheduled PI gains obtained for the nominal model and 100 models generated by randomly perturbing the nominal model parameters. The shaded regions indicate the ranges bounded by the minimum and maximum gain values obtained at each vehicle speed from these perturbed models.
}
As shown in Fig.~\ref{fig:SiimResult_pure_GS_VRFT_gain_characteristics}, the proportional gain of the conventional controller becomes negative after the vehicle velocity enters the untrained region. 
In addition, the integral gain increases considerably as the vehicle velocity increases. 
Since the conventional VRFT cost function is constructed only from the measured
data obtained in the training velocity region, the gain-scheduling law
cannot be directly evaluated in the untrained region. Consequently,
polynomial extrapolation of the gain-scheduling law produces large gain
variations, leading to excessive steering inputs and deterioration of
the yaw-rate tracking performance. 
On the other hand, as shown in Fig.~\ref{fig:SiimResult_proposed_GS_VRFT_gain_characteristics}, the PI gains obtained using the proposed method remain within relatively
small ranges over the entire simulation period. 
Compared with the conventional method, the proposed method suppresses the excessive
variation in the scheduled gains outside the training region. 
This is because the yaw-rate response predicted by the physics-based vehicle
model is included in the parameter-tuning cost function for both the
training and untrained velocity regions. 
We also confirmed that the proposed method achieved good performance under a noisy condition, in which white noise with a power of \(1 \times 10^{-7}\) was added to the output. 
The resulting MSEs for the tracking performance in the conventional and the proposed methods were \(2.152 \times 10^{-1}\) and \(1.021 \times 10^{-4}\), respectively.

These results demonstrate that the pure direct data-driven gain-scheduled PI
controller can provide satisfactory tracking performance within the
training velocity region but cannot maintain its performance when
extrapolated to an untrained velocity region. In contrast, the proposed
physics-informed direct data-driven gain-scheduled PI controller achieves satisfactory
tracking performance over the entire velocity range while suppressing
excessive variations in the scheduled gains and steering input.

\section{CONCLUSION}
\label{sec:conclusion}
This paper proposed a physics-informed direct data-driven control (PI-D3C) for vehicle lateral control. Specifically, the  physics-informed  VRFT-based gain-scheduled PI controller has been introduced. In the pure data-driven VRFT-based approach, the gain-scheduled controller parameters are determined only from the measured input-output data. Therefore, although satisfactory control performance can be achieved within the velocity region represented in the training data, the controller performance may deteriorate when the gain-scheduling law is extrapolated to untrained velocity regions. To address this problem, the proposed method incorporates the yaw-rate response predicted by a physics-based vehicle dynamics model into the controller-tuning cost function. This enables the gain-scheduled controller to be designed not only for the velocity region covered by the measurement data but also over velocity regions that are insufficiently represented in the training data.

The effectiveness of the proposed method was evaluated through vehicle dynamics simulations. The pure direct data-driven gain-scheduled PI controller achieved satisfactory tracking performance within the training velocity range up to 30~km/h. However, in the untrained velocity region from 30 to 80 km/h, both the proportional and integral gains exhibited excessive variations, resulting in significant deterioration of the yaw-rate tracking performance. In contrast, the proposed physics-informed direct data-driven gain-scheduled PI controller maintained satisfactory reference model tracking performance over the entire velocity range. These results demonstrate that incorporating prior physical knowledge into VRFT is effective in improving the extrapolation performance of a gain-scheduled controller designed from limited input-output data.

Future work will focus on experimental validation of the proposed method.

\bibliographystyle{IEEEtran}
\bibliography{IEEEabrv,reference}\ 

\end{document}